\documentclass[prd,aps,preprint,showpacs]{revtex4}
\usepackage{epsf}
\usepackage{pslatex}
\def\mpi2{m_\pi^2}
\def\mK2{m_K^2}

\newcommand{\bea}{\begin{eqnarray}}
\newcommand{\eea}{\end{eqnarray}}
\newcommand{\be}{\begin{equation}}
\newcommand{\ee}{\end{equation}}
\newcommand{\nn}{\nonumber}
\newcommand{\VEV}[1]{\left\langle #1\right\rangle}
\def\maths#1{$#1$}

\begin{document}
\bibliographystyle{apsrev}
\epsfclipon

%%%%%%%%%%%%%%%%%%%%%%%%%%%%%% MACROS %%%%%%%%%%%%%%%%%%%%%%%%%%

\newcommand{\pbp}{\langle \bar \psi \psi \rangle}
\newcommand{\pbdmdup}{\left\langle \bar \psi \frac{dM}{du_0} \psi  
\right\rangle}

%%%%%%%%%%%%%%%%%%%%%%%%%%%%%% TITLEPAGE %%%%%%%%%%%%%%%%%%%%%%%%%%

% \draft command makes pacs numbers print
% \draft

% \preprint{}%,% }

\title{QCD equation of state with 2+1 flavors of improved  
staggered quarks}

\author{C. Bernard}
\affiliation{Department of Physics, Washington University, St.~Louis,  
MO 63130, USA}

\author{T. Burch}
\affiliation{Institut f\"ur Theoretische Physik, Universit\"at  
Regensburg, D-93040 Regensburg, Germany}

\author{C. DeTar} \affiliation{Physics Department, University of Utah,
Salt Lake City, UT 84112, USA}

\author{Steven Gottlieb and L. Levkova\footnote{Current address:  
Department of Physics,
University of Utah, Salt Lake City, UT 84112, USA}}
\affiliation{Department of Physics, Indiana University, Bloomington,  
IN 47405, USA}

\author{U.M. Heller}
\affiliation{American Physical Society, One Research Road, Box 9000,  
Ridge, NY 11961-9000, USA}

\author{J.E. Hetrick}
\affiliation{Physics Department, University of the Pacific, Stockton, CA 95211, USA}

\author{R. Sugar}
\affiliation{Department of Physics, University of California, Santa  
Barbara, CA 93106, USA}

\author{D. Toussaint}
\affiliation{Department of Physics, University of Arizona, Tucson, AZ  
85721, USA}
\date{\today}

\begin{abstract}
We report results for the interaction measure, pressure and energy  
density
for nonzero temperature QCD with 2+1 flavors of
improved staggered quarks. In our simulations we use a Symanzik improved
gauge action and the Asqtad $O(a^2)$ improved staggered quark action
for lattices with temporal extent $N_t=4$ and 6. The heavy quark mass  
$m_s$
is fixed at approximately  the physical strange quark mass and the
two degenerate light quarks have masses $m_{ud}\approx0.1\, m_s$ or  
$0.2\, m_s$.
The calculation of the thermodynamic observables employs the integral
method where energy density and pressure are obtained by integration
over the interaction measure.
\end{abstract}


\maketitle

\newpage

%%%%%%%%%%%%%%%%%%%%%%%%%%%%%% INTRODUCTION %%%%%%%%%%%%%%%%%%%%%%%%%%

\section{Introduction}
\label{sec:intro}

Ordinary hadronic matter undergoes a qualitative change into a quark-gluon 
plasma (QGP) at high
temperatures and/or densities. The QGP is a new state of strongly  
interacting matter
in which the basic constituents, quarks and gluons, are ``freed''  
from the color confinement of low temperature hadrons.
The phenomenon of color confinement is attributed to the non-perturbative 
structure of the QCD
vacuum at zero temperature. At high temperatures (and/or densities) this picture is modified to  
allow a deconfining transition.
However, the character of the QGP at temperatures up to at least  
several times the
transition temperature ($T_c\approx 170$ MeV) remains non-perturbative,
since in this temperature range
the strong coupling constant is still of $O(1)$, and the fundamental degrees  
of freedom
are more complex than simply free quarks and gluons.
Currently lattice QCD is the only theoretical tool that is suitable  
for tackling this inherently
strongly coupled system from first principles.

The QGP is studied experimentally in heavy-ion collisions at RHIC and  
CERN, in which the accessible
temperature range is up to about $3T_c$ \cite{Adcox:2005}. The data  
from these
experiments are mostly interpreted through hydrodynamical models \cite 
{Huovinen:2006jp}, which take the equation of state (EOS) of
the low- and high-temperature phases as essential inputs.
The hydrodynamic models
that include the QGP as the high-temperature phase use an ideal gas  
EOS for quarks and gluons
which, considering the temperature range, is bound to be an  
unsatisfactory approximation, perhaps
accounting in part for discrepancies between some of the current  
predictions and the experimental data.
This difficulty can be addressed by a realistic lattice QCD calculation of the 
EOS to serve as input for the hydrodynamics equations. 

The importance of
a realistic EOS of the QGP is not limited to the heavy-ion  
experiments. The EOS is also relevant to cosmology,
since it is believed that the QGP existed microseconds after the big  
bang. For example, the relic density of
weakly interacting massive particles is sensitive to the EOS of the  
QGP at these early stages
of the formation of the Universe~\cite{Hindmarsh:2005ix}. Another  
area of potential application of the EOS
is in the study of phenomena in the interior of dense neutron stars,  
where again the QGP is likely to exist.

The determination of the EOS through numerical simulation of lattice
QCD is challenging, since it requires a precise determination of
differences between high and low temperature quantities that have
inherent ultraviolet divergences.  Thus the most extensive simulations
to date are carried out on rather coarse lattices ($N_t = 4$)
\cite{Blum:1994zf,Karsch:2000ps}.  Improving the gauge and fermion
actions \cite{Engels:1996ag,Bernard:2005mf} helps reduce lattice
artifacts as does decreasing the lattice spacing ($N_t = 6$)
\cite{Bernard:2005mf,Aoki:2005vt}.  It is also important to
carry out simulations with a realistic light quark spectrum
\cite{Karsch:2000ps,Bernard:2005mf}.

In this paper, we report results of a simulation of the QCD EOS
at $N_t = 6$ with $2+1$ light flavors of $O(a^2)$
tadpole-improved (Asqtad) staggered quarks. The gauge action we use is a Symanzik
$O(a^2)$  
tadpole-improved one as well. 
Preliminary accounts were given at the Lattice 2005 and 2006  
conferences \cite{Bernard:2005mf, lat2006}.
The inclusion of the strange quark is of interest
to the phenomenological studies of the QGP since it can change the  
order of the phase transition
and influences strangeness production in the heavy-ion  
experiments. To determine the EOS
we use the integral method where the pressure and the energy density  
are calculated through
an integration over the interaction measure \cite{Engels:1990vr}.
The paths of the integration in the bare parameter space
 are approximately trajectories of constant physics.
 Along a trajectory of constant physics  
the heavy quark
mass ($m_s$) would be fixed to the physical strange quark mass and  
the $m_\pi/m_\rho$ ratio would be
kept constant. We approximate two such trajectories for which $m_\pi/ 
m_\rho\approx 0.3$
and 0.4, which correspond to light quark masses $m_{ud}\approx 0.1\, m_s$ and  
$0.2\, m_s$, respectively. Our calculations
are performed at $N_t=6$ for both trajectories, and we have an  
additional $N_t=4$ result for
the $m_{ud}\approx 0.1\, m_s$ trajectory. In this work we compare the  
EOS obtained using, first, the data from
the two different trajectories and, second, from the data with  
different $N_t$. We 
find that the differences are small in both cases.

%%%%%%%%%%%%%%%%%%%%%%%%%%%%  Section %%%%%%%%%%%%%%%%%%%%%%%%%%%%%%%%

\section{The Integral Method for the EOS Determination}
\label{sec:analytic}

In this section we give a brief description of the formalism of the  
integral method as applied
to the specific improved actions that we use. The analytic form of  
the EOS is derived from the
following thermodynamics identities
\be%gin{equation}
   \varepsilon V = - \left.\frac{\partial \ln Z}{\partial(1/T)}\right| 
_V,\hspace{0.7cm}
   \frac{p}{T} = \left.\frac{\partial \ln Z}{\partial V}\right|_T  
\approx \frac{\ln Z}{V},\hspace{0.7cm}
    I = \varepsilon - 3p = -\frac{T}{V} \frac{d \ln Z}{d \ln a},
\label{eq:ids}
\end{equation}
where $\varepsilon$ is the energy density, $p$ is the pressure and $I 
$ is the interaction measure.
The spatial volume is $V=N_s^3a^3$ for lattice spacing $a$, and the  
temperature is $T=1/(N_ta)$. The derivative of the
partition function $Z$ with respect to the logarithm of the lattice  
spacing, $\ln a$, should be understood
as taken along a trajectory of constant physics. In the explicit form of  
the partition function
\be
   Z = \int dU \exp\left\{-S_g + \sum_f (n_f/4) {\rm Tr}\ln[M 
(am_f,U,u_0)]\right\}
\end{equation}
the gauge action is given by
$ S_g = S_{\rm pl} + S_{\rm rt} + S_{\rm pg}$, with
\begin{eqnarray*}
   S_{\rm pl} &=& \beta \sum_{x,\mu<\nu} (1 - P_{\mu\nu}) \\
   S_{\rm rt} &=& \beta_{\rm rt} \sum_{x,\mu<\nu} (1 - R_{\mu\nu}) \\
   S_{\rm pg} &=& \beta_{\rm pg} \sum_{x,\mu<\nu<\sigma} (1 - C_{\mu 
\nu\sigma}).
\end{eqnarray*}
The real parts of the traces of the $1\times1$ plaquette  -- $P_{\mu\nu}$,  the $1\times2$ and $2\times1$ rectangle sum
-- $R_{\mu\nu}$, and the $1\times 1\times 1$ parallelogram -- $C_{\mu\nu 
\sigma}$, are
all divided by the number of colors. The gauge couplings in the above are  
defined as
\begin{eqnarray}
   \beta &=& 10/g^2 \nonumber \\
   \beta_{\rm rt} &=& -\frac{\beta}{20u_0^2} (1 + 0.4805  
\alpha_s) \\
   \beta_{\rm pg} &=& -\frac{\beta}{u_0^2} 0.03325 \alpha_s  
\nonumber
\label{eq:betas}
\end{eqnarray}
for $\alpha_s = -4 \ln(u_0)/3.0684$ and $u_0 = \VEV{P}^{1/4}$. The  
fermion matrix $M(am_f,U,u_0)$
corresponds to the Asqtad staggered action for a specific flavor $f$.

Using the identities in Eq.~(\ref{eq:ids}) and the explicit form of $Z 
$ we obtain the EOS expressions
\bea
\label{eq:I}
Ia^4&=&-6\frac{d \beta} {d \ln a}\Delta\VEV{P}
        - 12\frac{d \beta_{\rm rt}} {d \ln a}\Delta\VEV{R}
        - 16\frac{d \beta_{\rm pg}} {d \ln a}\Delta\VEV{C}\\
     &-& \sum_f \frac{n_f}{4}\left[\frac{d (m_f a)}{d \ln a}
            \Delta\VEV{\bar \psi \psi}_f
        + \frac{d u_0}{d \ln a}
        \Delta\VEV{\bar\psi\, \frac{d M}{d u_0}\, \psi}_f\right]\nn\\
\label{eq:p}
   pa^4 &=& -\int_{\ln a_0}^{\ln a} I(a^\prime)(a^\prime)^4
          \, d\ln a^\prime \\
\label{eq:e}
  \varepsilon a^4 &=& (I+3p)a^4.
\eea
The various fermionic and gluonic observables in the EOS are  
calculated at nonzero temperature (fixed $N_t < N_s$)
and on zero-temperature lattices ( $N_t \geq N_s$). The symbol $\Delta 
$ in the EOS expressions denotes
their nonzero and zero-temperature differences. All measurements are  
taken along a
trajectory of constant physics, which we parameterize
with the lattice spacing $a$. The couplings $\beta$, $\beta_ 
{\rm rt}$, $\beta_{\rm pg}$, masses
$am_f$ and tadpole factor $u_0$ are all functions of $\ln a$ along
this trajectory. We use these functions to determine the derivatives  
of the bare parameters
with respect to $\ln a$ as needed for the EOS. The lower integration  
limit, $\ln a_0$, in Eq.~(\ref{eq:p})
should be taken at a coarse enough lattice spacing that the pressure difference
is negligible.

%%%%%%%%%%%%%%%%%%%%%%%%%%%%  Section %%%%%%%%%%%%%%%%%%%%%%%%%%%%%%%%

\section{Simulations on Trajectories of Constant Physics}
\label{sec:scales}

As already mentioned in the previous sections, for our simulations we  
use the one-loop Symanzik improved (L\"uscher-Weisz)
gauge action and the Asqtad quark action
\cite{Orginos:1998ue,Toussaint:1998sa,Lepage:1998vj}.
Both actions are tadpole improved and the leading
discretization errors are $O(a^2\alpha_s,a^4)$. There are many  
features of the Asqtad action that make it
well suited for high temperature studies. It has excellent scaling  
properties
leading to faster convergence to the continuum limit and the  
dispersion relations
for free quarks are much better than the ones for the standard Wilson  
or staggered actions.
Another very important property of the Asqtad action is the much  
reduced taste symmetry breaking
compared with the conventional staggered action.
All this translates into decreased lattice artifacts above the phase  
transition.

\begin{figure}[h]
\epsfxsize=12cm
\begin{center}
\epsfbox{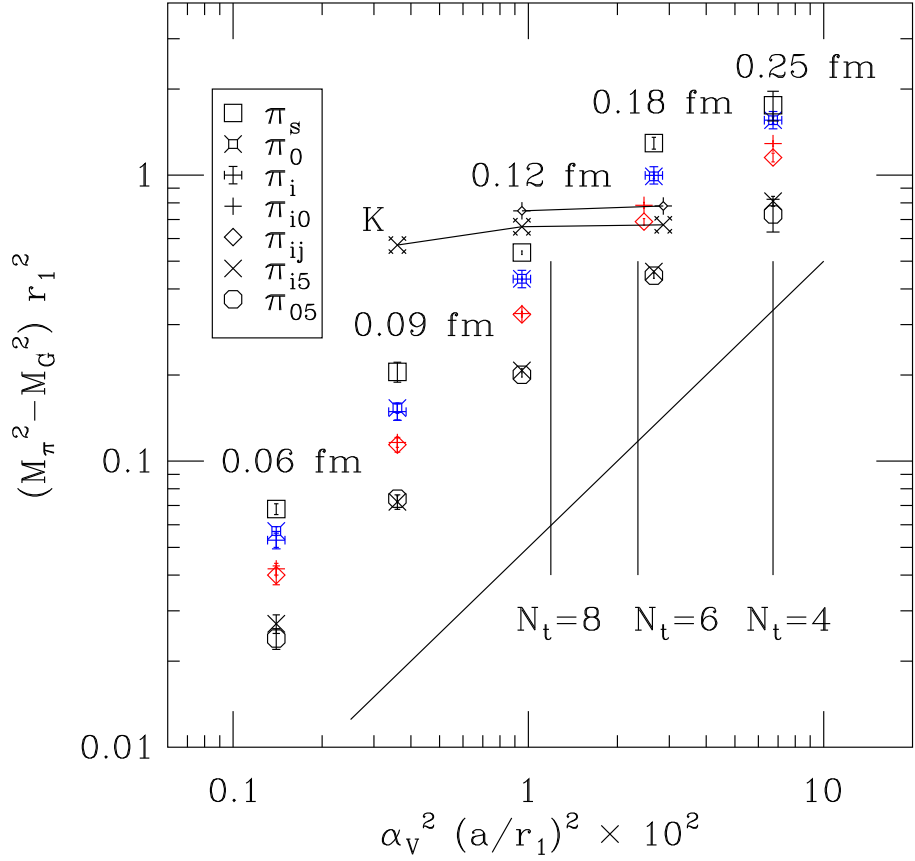}
\end{center}
\caption{Pion taste splitting relative to the Goldstone pion mass in
  units of $r_1 =0.318\, (7)\, (4)$~fm {\it vs.} the lattice scaling
  variable $(a/r_1)^2 \alpha_V(a)^2$ in a log-log plot. Here
  we take $\alpha_V(a)= 12 \pi/ [54 \ln[(3.33/a \Lambda)]$ with
  $\Lambda = 319$  MeV. The rising line has slope 1.
  The fancy diamonds locate the kaon splittings $(m_K^2 - m_G^2)r_1^2$
  for $m_{ud}\approx 0.2\, m_s$.  The fancy crosses do the same for
  $m_ {ud}\approx 0.1\, m_s$.  The vertical lines indicate the
  approximate lattice spacing at the crossover temperature for various
  $N_t$. Data are from \cite{Aubin:2004wf} and unpublished
  simulation results. The pion taste assignments are given in the
  gamma matrix basis. The taste singlet is denoted $\pi_s$.
}
\label{fig:split}
\end{figure}

It would seem important for studying the strange quark physics of the
plasma that the kaon mass be heavier than the pion.  In the staggered
fermion scheme each meson state appears in a taste multiplet of 16.
With improvement of the fermion action the splittings are considerably
reduced.  The splittings in meson mass squared are expected to vanish
in the continuum limit as $a^2 \alpha_V^2$.  Shown in a log-log plot
in Fig.~\ref{fig:split} are pion taste splittings relative to the
Goldstone pion mass for five lattice spacings.  The solid line shows
the expected scaling slope.  The trend is consistent with the scaling
prediction.  Shown also are splittings of the lowest member of the
kaon multiplet, relative to the Goldstone pion mass for the two
choices of $m_{ud}/m_s$ in the thermodynamics study.  The vertical
lines locate the lattice spacing at the crossover temperature $T_c$
(about 190 MeV for our unphysical light quark masses) for various
$N_t$.  Note that the temperature then increases as we move to the
left.  Our nonzero temperature studies are at $N_t=4$ and 6. At $N_t =
4$ the lightest kaon at $T_c$ has approximately the same mass as the
lowest non-Goldstone pion.  As the figure shows, at $N_t = 4$ the kaon
and pion taste multiplets are non-overlapping at approximately $T \ge 2
T_c$. At the $N_t = 6$ crossover the situation has improved and the
multiplets are non-overlapping at approximately $T \ge 4 T_c/3$.
Clearly, $N_t = 8$ would be even better for this action.

At
$N_t = 6$ the taste-splitting is about half as large as at $N_t=4$. 
One of our goals was to determine
to what extent the increase in $N_t$ from 4 to 6 influences the EOS.

\begin{figure}[ht]
\epsfxsize=13cm
\begin{center}
\epsfbox{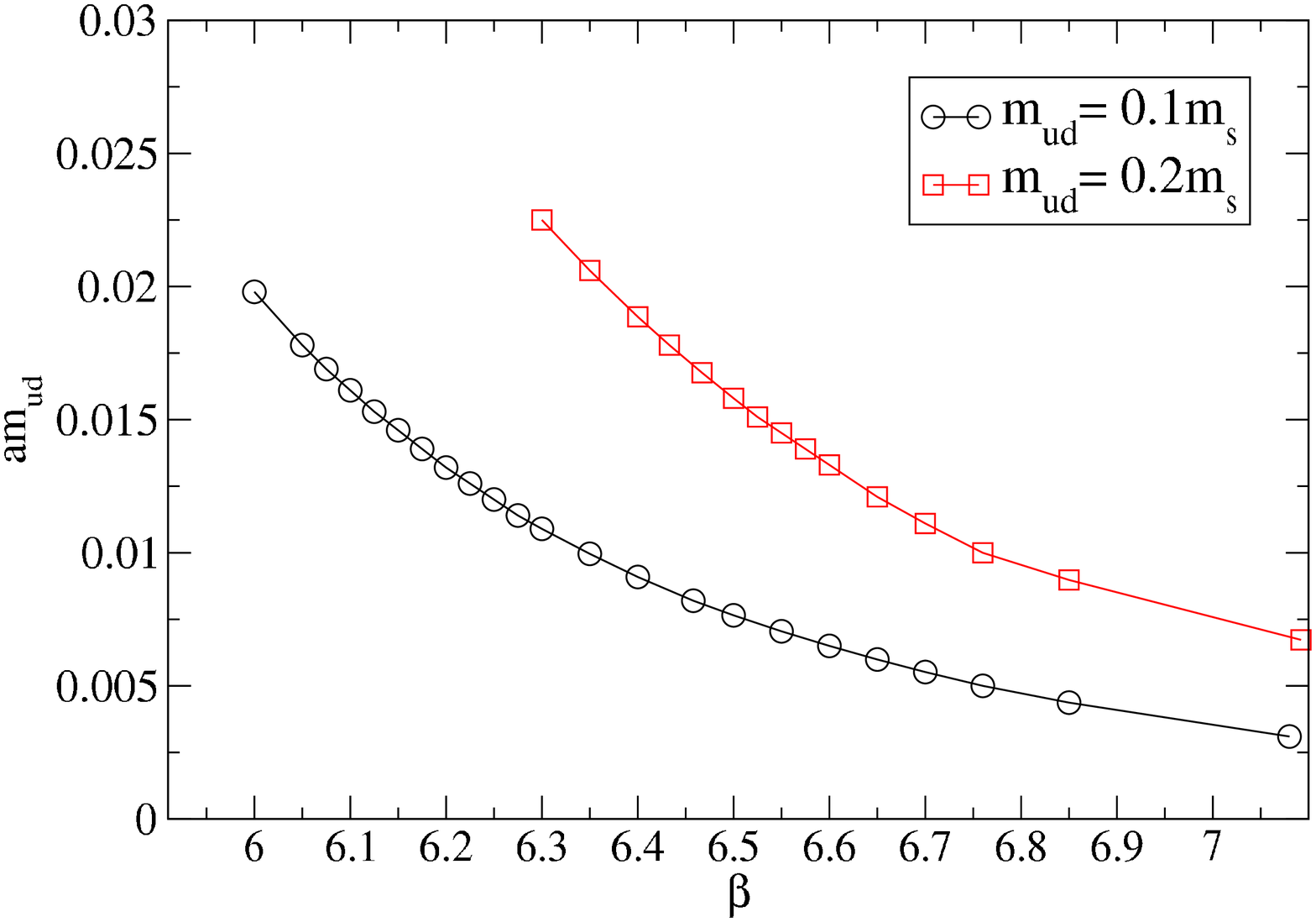}
\end{center}
\caption{Plot of the two trajectories of constant physics in  
the $(am_{ud}, \beta)$ plane.
}
\label{fig:traj}
\end{figure}

In our simulations we use the dynamical $R$ algorithm \cite{Gottlieb:1987mq}
with step size  
equal to
the minimum of 0.02 and $2\, am_{ud}/3$. For some runs the step size was  
chosen  to be even smaller.
Our aim is to generate zero- and nonzero-temperature ensembles of  
lattices with
action parameters chosen so that a trajectory of constant physics ($m_ 
\pi/m_\rho = {\rm const}$)
is approximated. Along the trajectory the heavy quark mass is tuned  
to
the strange quark mass within 20\%.
We work with two such trajectories: $m_{ud}\approx 0.2\, m_s$ ($m_\pi/m_ 
\rho \approx 0.4$) and
$m_{ud}\approx 0.1\, m_s$ ($m_\pi/m_\rho \approx 0.3$) as shown in Fig.~ 
\ref{fig:traj}.

The construction of each trajectory begins
       with ``anchor points" in $\beta$, where the hadron spectrum has
       been previously studied and the lattice strange quark mass
       has been tuned to approximate the correct strange hadron spectrum
       \cite{Aubin:2004wf}.
       We adjusted the value of $a m_{ud}$ at the anchor points to  
give a
       constant (unphysical) ratio $m_\pi/m_\rho$.  Between these points
       the trajectory is then interpolated, using a one-loop
       renormalization-group-inspired formula.  That is, we interpolate
       $\ln(a m_s)$ and $\ln(a m_{ud})$ linearly in $\beta$.  Since  
we have
       three anchor points for the $m_{ud}\approx 0.2\, m_s$ trajectory,
       namely $\beta = 6.467$, 6.76, and 7.092, our interpolation is
       piecewise linear.  For the trajectory $m_{ud}\approx 0.1\, m_s$  
we use
       two anchor points at $\beta = 6.458$ and 6.76. Explicitly the  
parameterization of the
$m_{ud}\approx 0.2\, m_s$ trajectory is
\bea
\label{eq:0.2ms}
am_s &=&\left\{
\begin{tabular}{ll}
$0.082 \,\exp\left((\beta-6.4674)\frac{\ln(0.050/0.0820)} 
{(6.76-6.4674)}\right)$,&$\beta\in[6.467,6.76]$\\
$0.05 \, \exp\left((\beta - 6.76)\frac{\ln(0.031/0.05)}{(7.092-6.76)} 
\right)$,&$\beta\in[6.76,7.092]$\\
\end{tabular}
\right.\\
am_{ud} &=& \left\{
\begin{tabular}{ll}
$0.01675\,\exp\left((\beta-6.4674)\frac{\ln(0.010/0.01675)} 
{(6.76-6.4674)}\right),$&$\beta\in[6.467,6.76]$\\
$0.01\,  \exp\left((\beta - 6.76)\frac{\ln(0.00673/0.01)} 
{(7.092-6.76)}\right),$&$\beta\in[6.76,7.092].$\\
\end{tabular}
\right.
\eea
The parameterization of the $m_{ud}\approx 0.1\, m_s$ trajectory with $ 
\beta\in[6.458, 6.76]$ is
\bea
am_s &=& 0.05\exp\left((\beta-6.76)\frac{\ln(0.082/0.05)} 
{(6.458-6.76)}\right)\\
am_{ud} &=& 0.005\exp\left((\beta-6.76)\frac{\ln(0.0082/0.005)} 
{(6.458-6.76)}\right).
\label{eq:0.1ms}
\eea
For both trajectories, for values of $\beta$ out of the interpolation  
intervals, the
parameterization formulae are used to perform extrapolations.
The run parameters of the two trajectories at different $N_t$
are summarized in Tables~\ref{tab:run0.1m_s4}, 
\ref{tab:run0.1m_s} and \ref{tab:run0.2m_s}.
It is apparent from Eqs.~(7) and (8) that values of the quark mass
ratio $m_{ud}/m_s$ in Table~\ref{tab:run0.2m_s} deviate slightly from 0.2,
since it was our initial
intention to keep the hadron masses constant instead.  In subsequent
practice, as in the $0.1 m_s$ trajectory, we chose the more convenient
alternative  of keeping the quark mass ratio fixed.
\begin{table}[ht]
\begin{tabular}{llllccccc}
\hline\hline
  \maths{\beta} & \maths{am_{ud}} & \maths{am_s} & \maths{u_0} &  
\maths{V_{T\neq0}} & Trajectory &\maths{V_{T=0}}& Trajectory &\maths{a} [fm]\\
\hline
\maths{^\star}6.000 & 0.0198 & 0.1976 & 0.8250 &\maths{12^3\times4} & 
1800& \maths{12^4}&500&0.366\\
\maths{^\star}6.050 & 0.0178 & 0.1783 & 0.8282 &\maths{12^3\times4} & 
1800&\maths{12^4}&500&0.334\\
              6.075 & 0.0169 & 0.1695 & 0.8301 &\maths{12^3\times4}  
&1800& & &  0.319\\
\maths{^\star}6.100 & 0.0161 & 0.1611 & 0.8320 &\maths{12^3\times4} & 
1800&\maths{12^4}&500&0.306\\
              6.125 & 0.0153 & 0.1533 & 0.8338 &\maths{12^3\times4}  
&2800&&  &           0.293\\
\maths{^\star}6.150 & 0.0146 & 0.1458 & 0.8356 &\maths{12^3\times4} & 
3800&\maths{12^4}&500&0.281\\
              6.175 & 0.0139 & 0.1388 & 0.8374 &\maths{12^3\times4}  
&3800&&&             0.269\\
\maths{^\star}6.200 & 0.0132 & 0.1322 & 0.8391 &\maths{12^3\times4} & 
3800&\maths{12^4}&400&0.258\\
              6.225 & 0.0126 & 0.126 & 0.8407 &\maths{12^3\times4}  
&3800&&             &0.248\\
\maths{^\star}6.250 & 0.012 & 0.1201 & 0.8424 &\maths{12^3\times4} & 
3800&\maths{12^4}&500&0.238\\
\maths{^\star}6.275 & 0.0114 & 0.1145 & 0.8442 &\maths{12^3\times4}  
&2800& $12^4$  &500&   0.229\\
\maths{^\star}6.300 & 0.0109 & 0.1092 & 0.8459 &\maths{12^3\times4} & 
1800&\maths{12^4}&2100&0.220\\
\maths{^\star}6.350 & 0.00996 & 0.0996 & 0.8491 &\maths{12^3\times4} & 
1800&\maths{12^4}&2100&0.204\\
              6.400 & 0.00909 & 0.0909 & 0.8520 &\maths{12^3\times4}  
&1800&&             &0.190\\
\maths{^\star}6.458 & 0.0082 & 0.082 & 0.8549 &\maths{12^3\times4} & 
1800&\maths{12^4}&2100&0.175\\
              6.500 & 0.00765 & 0.0765 & 0.8570 &\maths{12^3\times4}  
&1800&&           &  0.165\\
\maths{^\star}6.550 & 0.00705 & 0.0705 & 0.8593 &\maths{12^3\times4} & 
1800&\maths{20^4}&2100&0.155\\
              6.600 & 0.0065 & 0.065 & 0.8616 &\maths{12^3\times4}  
&1800&&             &0.145\\
\maths{^\star}6.650 & 0.00599 & 0.0599 & 0.8636 &\maths{12^3\times4} & 
1800&\maths{24^4}&2100&0.137\\
              6.700 & 0.00552 & 0.0552 & 0.8657 &\maths{12^3\times4}  
&1800&&             &0.129\\
\maths{^\star}6.760 & 0.005 & 0.05 & 0.8678 &\maths{16^3\times4} & 
800&\maths{24^3\times 64}&2100&0.120\\
\maths{^\star}6.850&  0.00437& 0.0437& 0.8710& \maths{16^3\times4} &  
800&\maths{32^4}&300& 0.109\\
\maths{^\star}7.080& 0.0031& 0.031& 0.8779& \maths{16^3\times4} & 
800&\maths{40^3\times96} &1500& 0.086\\
\hline\hline
\end{tabular}
\caption{Run parameters of the trajectory with $m_{ud}\approx 0.1\, m_s$  
at $N_t=4$. The asterisk indicates parameter sets for which
both zero and nonzero temperature runs were performed. The columns labeled "Trajectory" indicate
the number of thermalized trajectories. Last column  
shows the lattice spacing as determined from Eq.~(\ref{eq:a}).}
\label{tab:run0.1m_s4}\vspace{3mm}
\end{table}

\begin{table}
\begin{tabular}{llllccccc}
\hline\hline
\maths{\beta} & \maths{am_{ud}} & \maths{am_s} & \maths{u_0} & \maths 
{V_{T\neq0}} & Trajectory &\maths{V_{T=0}}&Trajectory &\maths{a} [fm] \\
\hline
\maths{^\star}6.300 & 0.0109 & 0.1092 & 0.8459 &\maths{12^3\times6}  
&3100& \maths{12^4}&2100 & 0.220 \\
\maths{^\star}6.350 & 0.00996 & 0.0996 & 0.8491 &\maths{12^3\times6}  
&2900& \maths{12^4}&2100 & 0.204\\
         6.400 & 0.00909 & 0.0909 & 0.8520 &\maths{12^3\times6} &   
2900&&&0.190\\
\maths{^\star}6.458 & 0.0082 & 0.082 & 0.8549 &\maths{16^3\times6}  
&2140& \maths{12^4}&2100 &0.175\\
         6.500 & 0.00765 & 0.0765 & 0.8570 &\maths{12^3\times6} &   
2900&&&0.165\\
\maths{^\star}6.550 & 0.00705 & 0.0705 & 0.8593 &\maths{12^3\times6}  
&2900& \maths{20^4}&2100 &0.155\\
         6.600 & 0.0065 & 0.065 & 0.8616 &\maths{12^3\times6} &   
2900&&&0.145\\
\maths{^\star}6.650 & 0.00599 & 0.0599 & 0.8636 &\maths{12^3\times6}  
&2900& \maths{24^4} &2100&0.137\\
         6.700 & 0.00552 & 0.0552 & 0.8657 &\maths{12^3\times6} &   
2900&&&0.129\\
\maths{^\star}6.760 & 0.005 & 0.05 & 0.8678 &\maths{20^3\times6}  
&1000& \maths{24^3\times 64}&2100 &0.120\\
\maths{^\star}6.850&  0.00437& 0.0437& 0.8710& \maths{18^3\times6} &  
1300& \maths{32^4}&300& 0.109\\
\maths{^\star}7.080 & 0.0031 & 0.031 & 0.8779 &\maths{18^3\times6}  
&2200& \maths{40^3\times 96}&1500 &0.086\\
\hline\hline
\end{tabular}
\vspace{3mm}
\caption{ Same as Table~\ref{tab:run0.1m_s4} but for trajectory $m_{ud}\approx 0.1\, m_s$  
at $N_t=6$.
}
\label{tab:run0.1m_s}
\end{table}

\begin{table}[ht]
\begin{tabular}{llllccccc}
\hline\hline
\maths{\beta} & \maths{am_{ud}} & \maths{am_s} & \maths{u_0} & \maths 
{V_{T\neq0}} &Trajectory& \maths{V_{T=0}}&Trajectory &\maths{a} [fm] \\
\hline
\maths{^\star}6.300 & 0.0225 & 0.1089 & 0.8455 &  \maths{12^3\times6}  
&3000& \maths{12^4}&2100& 0.224\\
\maths{^\star}6.350 & 0.0206 & 0.1001 &0.8486&\maths{12^3\times6}& 
3000&\maths{12^4}&2100& 0.208 \\
  6.400 &0.01886 &0.0919 & 0.8512&\maths{12^3\times6}&3000&& & 0.193 \\
  6.433 &0.0178 &0.087 &0.8530&\maths{12^3\times6}&800&& & 0.184\\
\maths{^\star}6.467 & 0.01676 &0.0821 &0.8549&\maths{16^3\times6}& 
2200&\maths{16^3\times48}&1225& 0.176 \\
  6.500 &0.0158 &0.0776 &0.8568& \maths{12^3\times6}&3000&&& 0.168\\
\maths{^\star}6.525 &0.0151 &0.0744 &0.8580&\maths{12^3\times6}&  
3000&\maths{12^4}&2100& 0.162 \\
  6.550 &0.0145 &0.0713 &0.8592 &\maths{12^3\times6}&3000&&& 0.157\\
\maths{^\star}6.575 &0.0139 &0.0684 &0.8603 &\maths{12^3\times6}&  
3000&\maths{16^4}&1760& 0.152\\
  6.600 &0.0133 &0.0655 &0.8614&\maths{12^3\times6}&3000&&& 0.147 \\
\maths{^\star}6.650 &0.0121 &0.0602 &0.8634&\maths{12^3\times6}&  
3000&\maths{20^4}&836& 0.138\\
  6.700 &0.0111 &0.0553 &0.8655 &\maths{12^3\times6}&3100&&& 0.130\\
\maths{^\star}6.760 &0.01 &0.05 &0.8677& \maths{20^3\times6}& 
1935&\maths{20^3\times64}&825& 0.121\\
\maths{^\star}6.850 & 0.00898&  0.0439& 0.8710& $12^3\times6$ &3000& $24^3 
$&740&0.110 \\
  7.092 &0.00673 &0.031 &0.8781& \maths{12^3\times6}&3000&&& 0.086\\
  7.090 &0.0062 &0.031 &0.8782& & &\maths{28^3\times96}&565& \\
\hline\hline
\end{tabular}
\vspace{3mm}
\caption{Same as Table~\ref{tab:run0.1m_s4} but for trajectory $m_{ud}\approx 0.2m_s$  
at $N_t=6$. The last row is a run which does not lie on the trajectory and was 
used only for zero-temperature extrapolations.
}
\label{tab:run0.2m_s}
\end{table}

\begin{figure}[ht]
\epsfxsize=130mm
\begin{center}
\epsfbox{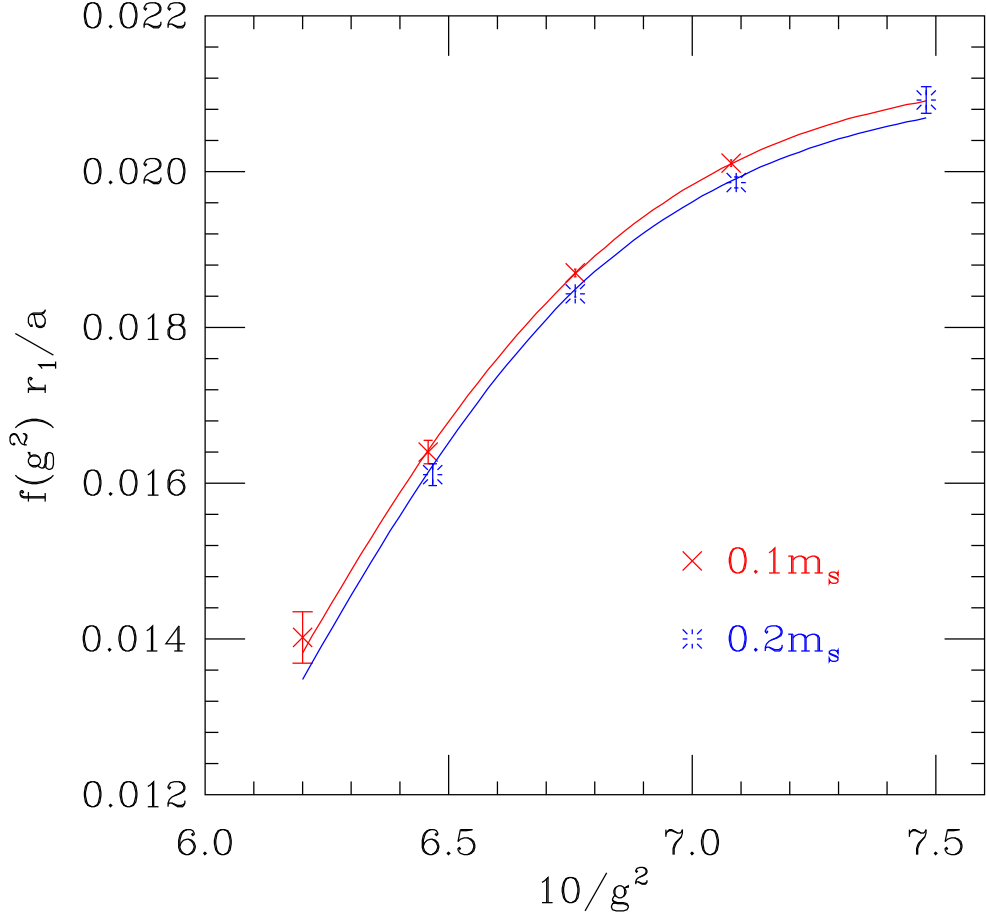}
\end{center}
\caption{Inverse lattice spacing in units of $f(g^2)r_1$ {\it vs.} gauge
coupling $\beta = 10/g^2$, based on the best fit parameterization
Eq.~(\ref{eq:a}).  The fitting function is evaluated along the two lines of
constant physics, namely $m_{ud} \approx 0.1 m_s$ and $0.2 m_s$.  It
is derived from forty measured values of $r_1/a$.  Eight of them lie
on the trajectories of constant physics and are plotted here.}
\label{fig:r1da}
\end{figure}

For the purpose of the EOS determination, the trajectories of constant physics  
are most conveniently parameterized
by the lattice spacing $a$, as discussed at the end of the previous  
section. To calculate the various
derivatives of the bare parameters with respect to $\ln a$ we need to  
determine the functional dependence
$\ln a(\beta)$.
The lattice spacing is determined using the method
of Ref.~\cite{Aubin:2004wf}.  On a large set of zero-temperature ensembles 
the static potential is measured to determine the modified Sommer
parameter $r_1$ \cite{Bernard:2000gd} in lattice units.  Specifically, $r_1$ is 
defined by $r_1^2F_{\bar{q}q}(r_1)=1$.  All available measurements
of $r_1/a$ are then fit to 
the following  
asymptotic-freedom-inspired form \cite{Allton:1996kr,Allton:1996dn}
\be
\frac{a}{r_1} = \frac{c_0f(g^2) + c_2g^2f^3(g^2) + c_4g^4f^3(g^2)}{1+d_2g^2f^2(g^2)}.
\label{eq:a}
\end{equation}
The definition of
\be
f(g^2)=(b_0g^2)^{-b_1/(2b_0^2)}e^{-1/(2b_0g^2)}
\end{equation}
involves the universal beta-function coefficients for
massless three-flavor QCD, $b_0$ and $b_1$. The coefficients $c_0$,  
$c_2$ and $c_4$ are
\bea
c_0 &=& c_{00} + (c_{01u}am_{ud}+ c_{01s}am_s)/f(g^2) + c_{02}(2am_{ud}+am_s)^2/f^2 
(g^2)\nn\\
c_2 &=& c_{20} + c_{21}(2am_{ud}+am_s)/f(g^2)\nn\\
c_4 &=& c_{40}\nn\\
d_2 &=& d_{20},\nn
\eea
where
$c_{00}   = 46.766(447)$,
$c_{01u}   = 0.526(122)$,
$c_{01s}   = 0.1817(708)$,
$c_{02}   = -0.00403(204)$,
$c_{20}   =  -4.702(175)\times10^5$,
$c_{21}   = 3.321(511)\times10^3$, 
$c_{40}   =   3.943(84)\times10^5$ and 
$d_{20}   = 1.276(484)\times10^3$.
The fit has $\chi^2/{\rm DoF}\approx1.3$ and a confidence level of  
approximately 0.13.  This parameterization provides a
determination of $r_1/a$ along our trajectories of constant physics (Fig.~\ref{fig:r1da}).  
Independently of the fit, the absolute scale for $a$ is set from a 
determination of the
$\Upsilon(2S-1S)$ splitting on a subset of these zero-temperature
ensembles \cite{Wingate:2003gm,Gray:2005ur}. An extrapolation
to zero lattice spacing then gives  $r_{1}=0.318\, (7)\, (4)$ fm
\cite{Aubin:2004wf}.  
This value was used in conjunction with the above parameterized value
of $r_1/a$ to define the physical lattice spacing in our simulations.

%%%%%%%%%%%%%%%%%%%%%%%%%%%%  Section %%%%%%%%%%%%%%%%%%%%%%%%%%%%%%%%

\section{Equation of State Results and Conclusions}
\label{sec:eos}

In the previous sections, we have outlined the method we follow to  
determine the temperature
dependence of the bulk thermodynamic quantities, namely, the 
interaction measure, pressure and energy density,
which constitute the EOS for the quark-gluon system. In this  
section we present our numerical results.

\begin{figure}[ht]
\epsfxsize=130mm
\begin{center}
\epsfbox{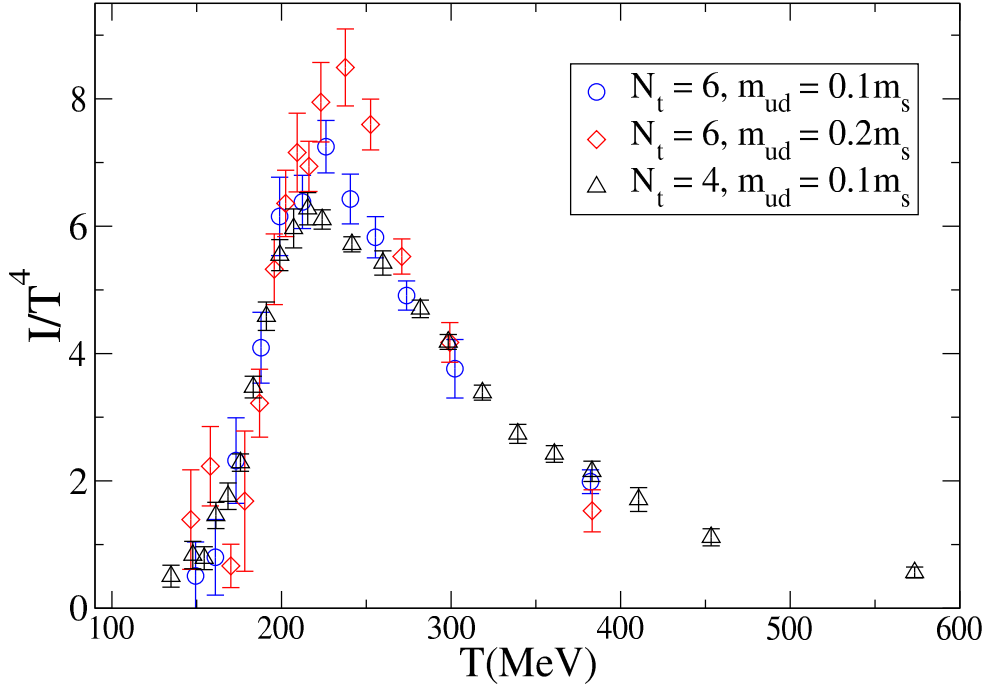}
\end{center}
\caption{The interaction measure is shown for both of the trajectories of
constant physics  
and the different $N_t$.}
\label{fig:I}
\end{figure}

According to the integral method, at the base of our calculation is  
the determination
of the interaction measure, which is straightforward from Eq.~(\ref 
{eq:I}).
The nonzero temperature value of the interaction measure needs to be  
corrected for the zero-temperature
contributions. This correction is done for about half of the runs by  
directly
measuring the zero-temperature values of the fermionic and gluonic  
observables involved in Eq.~(\ref{eq:I})
and subtracting their resultant zero-temperature contribution from  
the interaction measure at nonzero
temperature. For the rest of the runs, the zero-temperature correction  
is calculated by making
local interpolations. We need to determine as well the derivatives
$d \beta/d \ln a$,
$d \beta_{\rm rt}/d \ln a$,
$d \beta_{\rm pg}/d \ln a$,
$d (m_f a)/d \ln a$ and
$d u_0/d \ln a$ for each trajectory. For this purpose we take  
derivatives of
the $\ln(am_{ud})$ and $\ln(am_{s})$ trajectory parameterizations,  
using Eq.~(\ref{eq:0.2ms})~--~Eq.~(\ref{eq:0.1ms}),
polynomial fits to $u_0(\beta)$ for both trajectories, and the $a/r_1 
$ fit from Eq.~(\ref{eq:a}).
Figure~\ref{fig:I} shows the interaction measure as a function of the  
temperature
for both trajectories of constant physics and $N_t$'s.

\begin{figure}[ht]
\epsfxsize=130mm
\begin{center}
\epsfbox{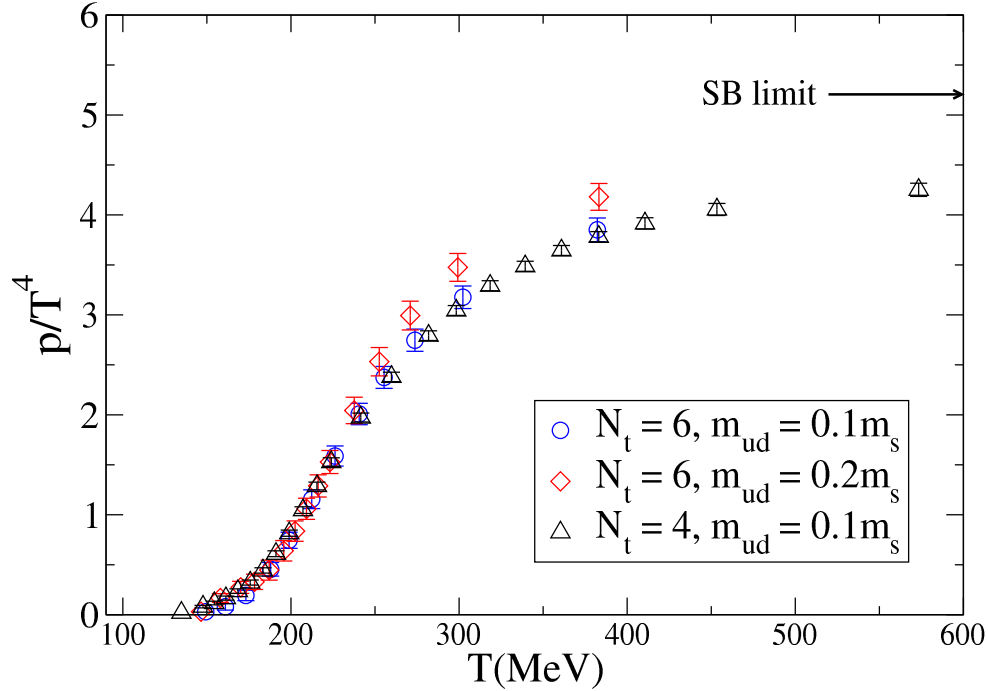}
\end{center}
\caption{The temperature dependence of the pressure for both of the  
trajectories of constant physics and the different $N_t$. The continuum Stefan-Boltzmann
 limit for 3 massless flavors is also shown.}
\label{fig:P}
\end{figure}

\begin{figure}[ht]
\epsfxsize=130mm
\begin{center}
\epsfbox{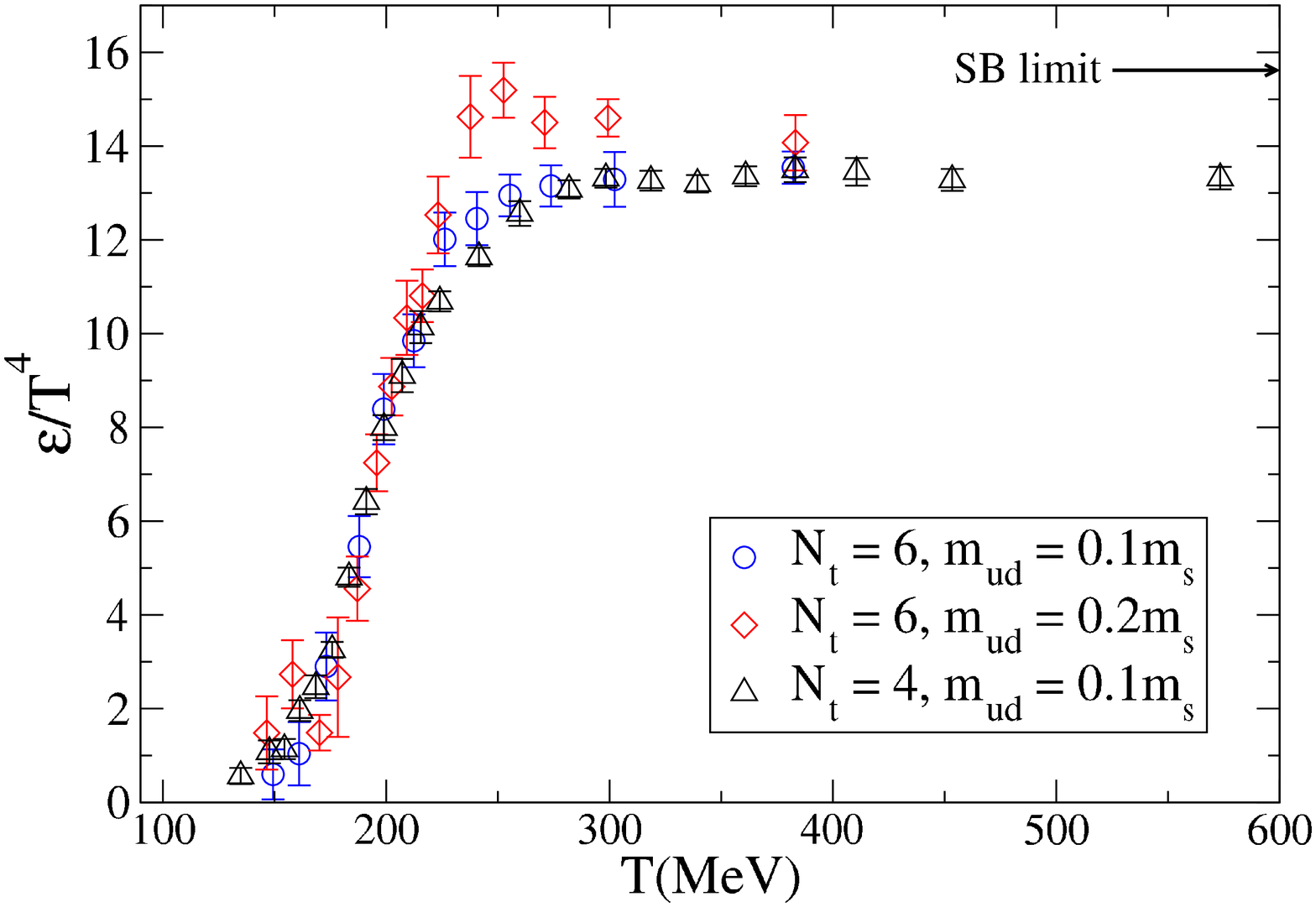}
\end{center}
\caption{The temperature dependence of the energy density for both of the trajectories of constant 
physics  and the different $N_t$.
To facilitate comparison with the ideal gas case, the continuum  
Stefan-Boltzmann limit for 3 massless flavors
is also shown.}
\label{fig:E}
\end{figure}

The pressure is obtained from the interaction measure by integration  
(Eq.~(\ref{eq:p})) using the trapezoid method and 
Fig.~\ref{fig:P} shows our results.  
Using both the interaction measure and the  
pressure, we calculate the energy density
(Eq.~(\ref{eq:e})).  The results are presented in Fig.~\ref{fig:E}.  
The statistical errors on all of the thermodynamic
quantities are calculated using the jackknife method, and
we ignore the insignificant errors on the derivatives of the bare  
parameters with respect to the lattice scale
mentioned above. The EOS data in the Figs.~\ref{fig:I} -- \ref 
{fig:E} is corrected
for the systematic errors due to the finite step size,
which are discussed later in this section and in more detail in the  
Appendix, and the choice of the lower integration limit in Eq.~(\ref{eq:p}).

From our results for the EOS we find that at the highest
studied temperature ($\sim 380$ MeV for $N_t=6$ and $\sim570$ MeV for  
$N_t=4$~) the energy density
is about 10 -- 15\% below the Stefan-Boltzmann three-flavor limit, which is  
evidence
that strong interactions between the plasma constituents 
persist in the high temperature
phase at several times $T_c$.
The comparison of the EOS for the two trajectories of constant physics  
at $N_t=6$ shows
some small differences. There is a difference in the interaction measure maxima, with  
the one from the
$m_{ud}\approx 0.2\, m_s$ trajectory somewhat larger than the $0.1\, m_s$  
trajectory one. Also, the pressure on the $m_{ud}\approx 0.2\, m_s$ trajectory at coarse
lattice spacing (see Fig.~\ref{fig:P}) is gradually becoming  
slightly larger with temperature 
than that on the $0.1\, m_s$ trajectory.  This result is contrary to
expectation. We consider that the cause is the accumulation of various systematic errors
in the pressure calculation which we discuss later in this section.
As a whole, the reduction of the mass of the degenerate light quarks from
$m_{ud}\approx 0.2\, m_s$ to $0.1\, m_s$ does not affect dramatically the basic
thermodynamic properties of the system.   
We see a lot of similarity between the EOS at
$N_t=4$ and 6 for the $m_{ud}\approx 0.1\, m_s$ trajectory. The main  
differences between the two available
$N_t$ results is again in the interaction measure,  where the maximum at 
$N_t=6$ is higher. 
 Although the discretization artifacts  
at $N_t=4$ are known to be
larger than in the $N_t=6$ case, we find that their effect on the EOS  
is not very pronounced.

\begin{figure}[ht]
\epsfxsize=130mm
\begin{center}
\epsfbox{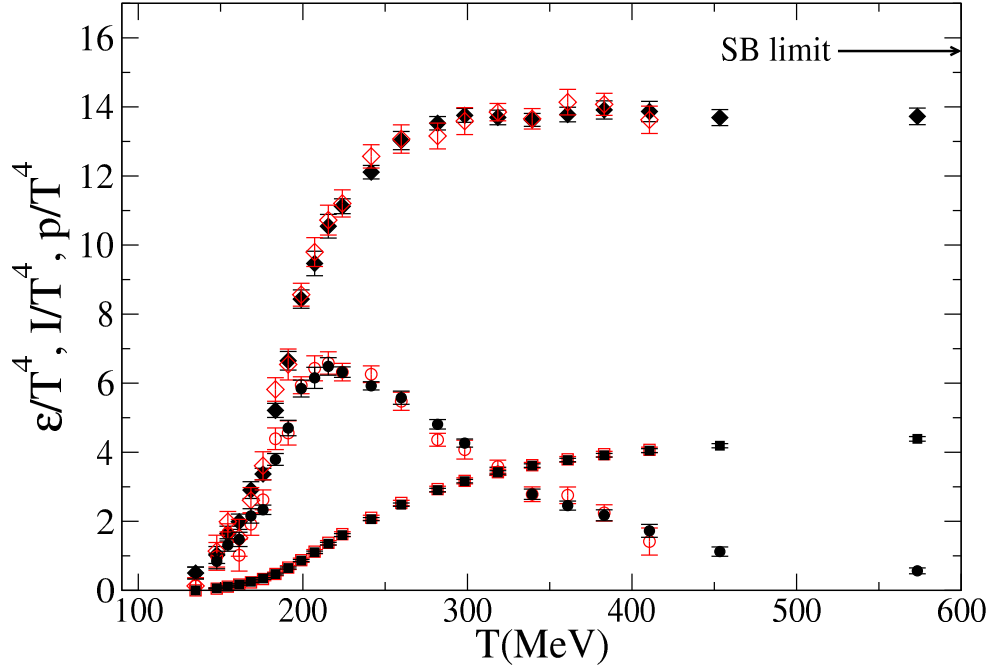}
\end{center}
\caption{
Volume dependence of results for $N_t=4$ with $m_{ud}\approx 0.1 m_s$. 
Empty
symbols are used for small volume ($V_s=8^3$) results, and filled symbols
are used for large volume ($V_s=12^3$ or $16^3$) results.  The energy
density,
pressure and interaction measure are plotted using diamonds, squares and
circles, respectively.  The data are not corrected for any systematic
errors.  We see no statistically significant volume dependence.
}
\label{fig:Vs}
\end{figure}

Our EOS calculation can be affected by the following systematic  
errors: finite volume effects,
finite step-size effects, the error in the determination of the  
lower integration limit in Eq.~(\ref{eq:p}), possible deviations from the
trajectories of constant physics  and the 
uncertainties in the scale determination from Eq.~(\ref{eq:a}).
First we discuss the scale determination error. The lattice spacing
is most difficult to obtain for the $N_t=4$ case in the low temperature 
region, where the lattices are coarse. We have estimated that a 5\%
error on the lattice scale there gives up to a three-sigma effect in 
the region of the interaction measure maximum.  
This translates as well into up to two-sigma effects on the energy density 
and pressure at high temperatures,
since errors accumulate in the integration needed to obtain these quantities.

To address the question of the finite volume effects we have  
conducted a set of runs at $N_t=4$ with
parameters from the $m_{ud}\approx 0.1\, m_s$ trajectory on lattices  
with smaller spatial volume -- $V_s=8^3$.
In Fig.~\ref{fig:Vs} the thermodynamic quantities calculated using  
$V_s=8^3$
are compared with the ones on the larger spatial volumes --  
$V_s=12^3, 16^3$. We find no statistically
significant difference which leads us to conclude that in our  
calculation
the finite volume effects are negligible.

The determination of the lower integration limit in Eq.~(\ref{eq:p})  
is potentially another source of
systematic error. The lowest available temperature in our  
calculations is around 135 MeV at $N_t=4$,
and 149 MeV at $N_t=6$. To estimate the pressure at these  
temperatures we calculate the
pressure of an ideal Bose gas of pions with masses similar to those  
in our simulations. The true
Goldstone pions on the physics trajectories have mass of $\sim 270$  
MeV and the rest of the
members of the pion multiplet are heavier. We estimated the heavy  
pion masses using extrapolations
of available data for the taste splitting in the pion multiplet,  
summarized in Fig.~\ref{fig:split}.
Including all of the pions according to their degeneracy, we estimate  
$p/T^4(T=135\,\, {\rm MeV}) \sim 0.02$ and
$p/T^4(T=149\, \,{\rm MeV}) \sim 0.03$ with about 30\% uncertainty 
in these values. 
Both of these estimations are comparable or a bit larger than the  
size of the statistical
error on the pressure at the closest available low temperatures.
Consequently we have corrected the pressure and energy density
by adding them to the data.
At high temperatures this correction is smaller than 
the statistical error.

The error due to deviations from the trajectories of constant
physics would be largest in the $N_t=4$ case, for
which the points around the transition region and at lower temperature
were obtained by extrapolations using Eqs.~(9) and (10). Indeed, a
later spectrum calculation near the transition, at $\beta=6.2$, showed
that
there is about a 10\% difference from the target value for $m_\pi/m_\rho$.
However, considering that the differences between the two $N_t=6$
trajectories, for which $m_\pi/m_\rho$ differs by about 30\%, is no more
than four sigma, we estimate the effect at about one
sigma in the transition region and smaller outside of it.
\begin{figure}[ht]
\epsfxsize=130mm
\begin{center}
\epsfbox{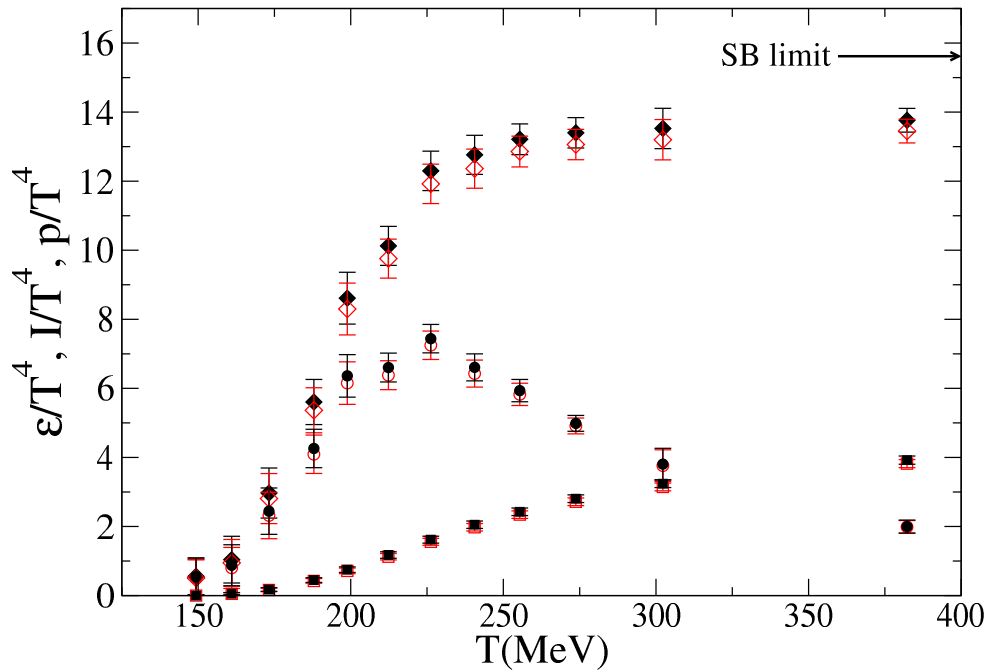}
\end{center}
\caption{
Effect of step-size corrections for $N_t=6$ with $m_{ud}\approx 0.1
m_s$.
We use filled (open) symbols to plot uncorrected (corrected) results.
The symbols for the energy density, pressure and interaction measure 
are diamonds, squares and circles, respectively. 
}
\label{fig:corr}
\end{figure}

The last potentially significant source of systematic error is the  
finite step
size used in the $R$ algorithm.  For $N_t = 4$ and 6 we have carried out
a set of test simulations at a larger step size in the $R$ algorithm to  
estimate their
effect. In addition we have performed some RHMC \cite{Clark:2004cp}  
calculations
to complement our finite step-size study.
Our analysis of the results is presented in the Appendix.  We
find that  the effect of the step-size corrections to the gauge  
observables on the
interaction measure is no larger than the size of our statistical error  
along the
$m_{ud}\approx 0.1\, m_s$ trajectory and negligible along  
the $m_{ud}\approx 0.2\, m_s$ one.  The
effects of the correction to the fermionic observables for both  
trajectories is small enough to be ignored.
We use the empirical formula in Eq.~(\ref{eq:stepsizeplaq}) with the  
parameters in
Eq.~(\ref{eq:coeff}) to compute the correction to the three gauge loop
observables for the $m_{ud}\approx 0.1\, m_s$ trajectory only. We do not  
correct the
$m_{ud}\approx 0.2\, m_s$ trajectory for finite step-size effects due to  
their smallness.
Figure~\ref{fig:corr} shows the EOS for the $m_{ud}\approx 0.1\, m_s$,
$N_t=6$ case with the finite step-size correction compared with the
uncorrected case.  The correction is no larger than our statistical
errors.  As explained in the appendix, we estimate the uncertainty in
the correction itself to be about 50\%.
\section*{ACKNOWLEDGMENTS}
We thank Frithjof~Karsch  and Peter~Petreczky for 
useful discussions on the scale determination.
Computations for this work were performed at Florida State
University, Fermi National Accelerator Laboratory (FNAL), Indiana University,
the National 
Center for Supercomputer Applications (NCSA), 
the National Energy Resources Supercomputer Center (NERSC),
the University of Utah (CHPC) and the University of California, Santa Barbara
(CNSI). 
This work was supported by the U.S. Department of Energy under contracts
DE-FG02-91ER-40628,      %C. Bernard
DE-FG02-91ER-40661      %S. Gottlieb
and
DE-FG03-95ER-40906, and       %D. Toussaint
National Science Foundation grants
PHY05-55243,             %C. DeTar
and
PHY04-56556.             %R. Sugar

%%%%%%%%%%%%%%%%%%%%%%%%%%%% REFERENCES %%%%%%%%%%%%%%%%%%%%%%%%%%%%%% 
%%%%

\bibliography{eos}

%%%%%%%%%%%%%%%%%%%%%%%%%%%%%% APPENDIX %%%%%%%%%%%%%%%%%%%%%%%%%%

\appendix
\section{Step-size dependence of EOS observables}

With the $R$ algorithm the integration step size
in the molecular dynamics evolution must be chosen small enough
to achieve the desired accuracy in observables of interest.  For most
%RS to achieve sufficient accuracy in the desired observables.  For most
practical purposes we have found errors at our standard small
production step sizes to be insignificant. (For a recent test see
Table IX of \cite{Aubin:2004wf}.) However, the observables required
for the equation of state must be measured to a very high accuracy,
since the small differences between the hot and cold measurements are
sensitive to even small systematic errors.  The most important
observable in this regard is the plaquette.  For the present study we
have developed a rough empirical method for estimating and correcting
for these errors in our simulations.

To estimate the step-size error within the $R$ algorithm requires
carrying out simulations at a range of step sizes and determining the
change in the observable as the step size tends to zero.  We have
carried out a number of such tests on hot and cold Asqtad lattices,
measuring most of the observables needed for the equation of state.
The RHMC algorithm, which we incorporated into our code at the end of
this study, does not suffer from such step size errors.  Thus for the
purpose of modeling step-size corrections we include results of some
RHMC calculations.  For our previous study of the equation of state
with the unimproved gauge action and naive staggered fermion action, we
made extensive measurements of the step-size dependence of the
plaquette and chiral condensate \cite{Bernard:1996cs}.

\begin{figure}[ht]
\begin{tabular}{ll}
  \epsfxsize=70mm
  \epsfbox{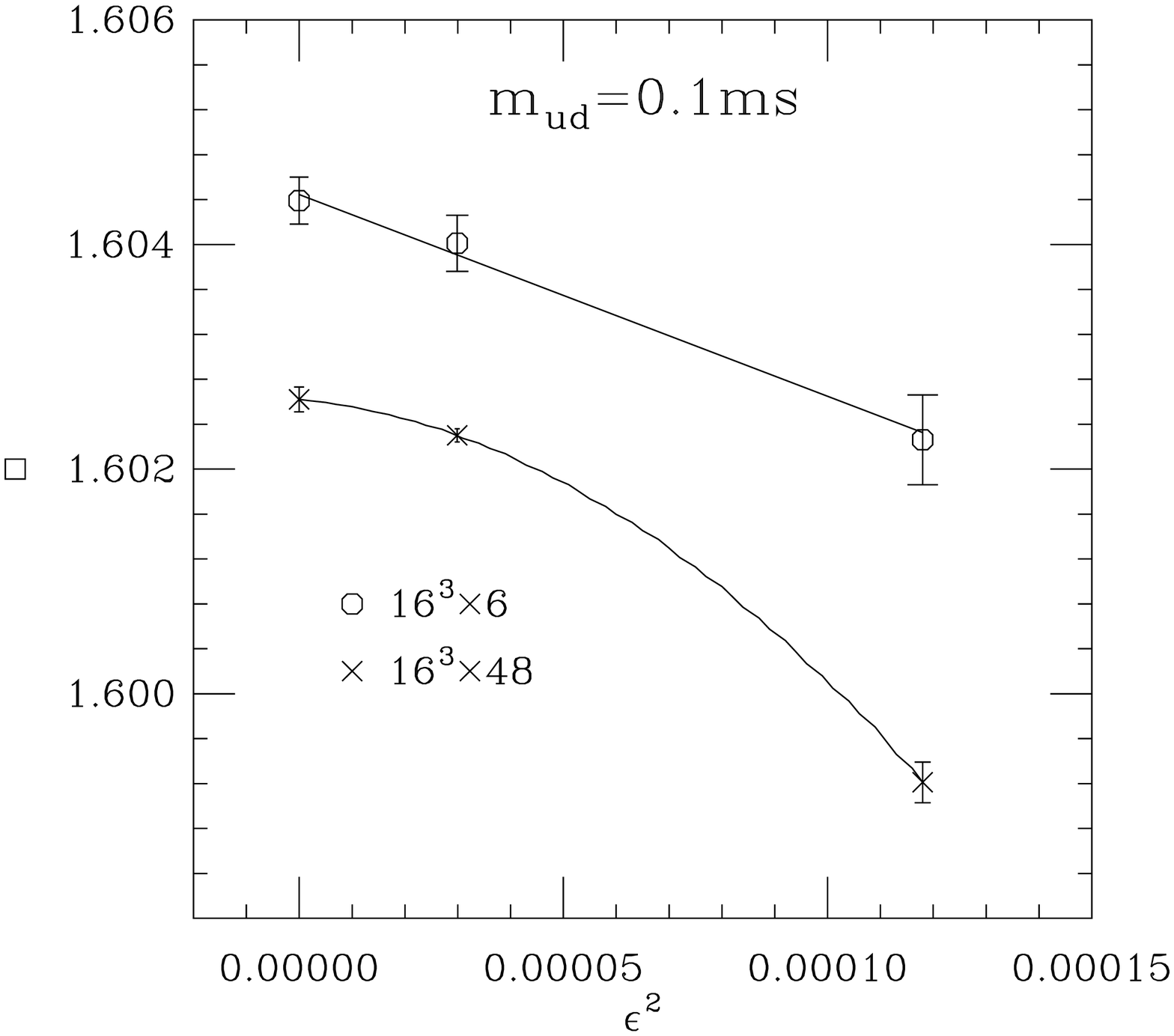}
&
  \epsfxsize=70mm
  \epsfbox{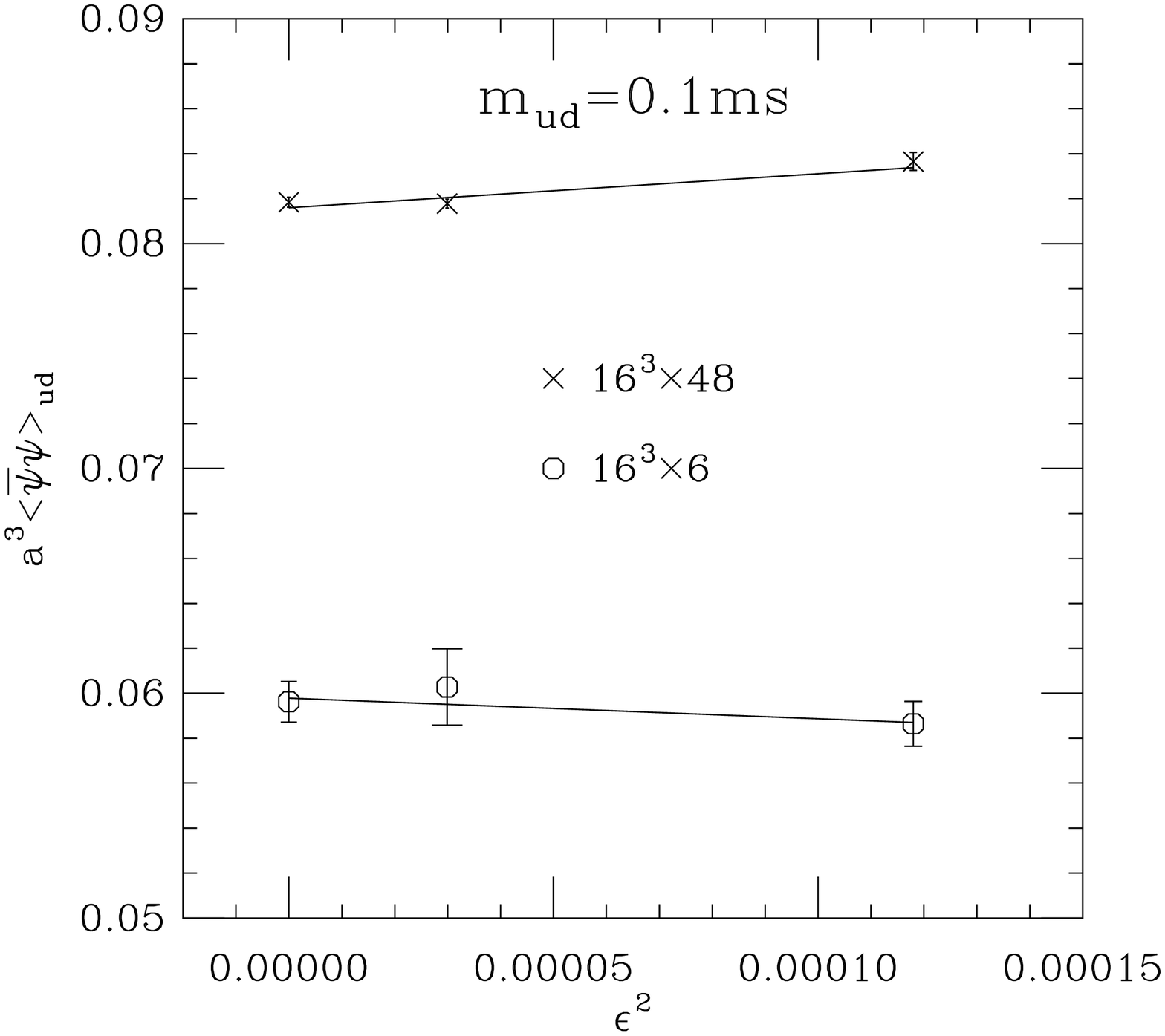}
\end{tabular}

\caption{Plaquette (left panel) and chiral condensate (right panel)
{\it vs} the squared step size $\epsilon^2$ for the improved action
for the ensemble at $\beta = 6.458$, $am_{ud} = 0.0082$ and $am_s =
0.082$.  The squared step size for the production of this ensemble is
0.00003 }
\label{fig:obsvseps0.1ms}
\end{figure}

\begin{figure}[ht]
\begin{tabular}{ll}
  \epsfxsize=70mm
  \epsfbox{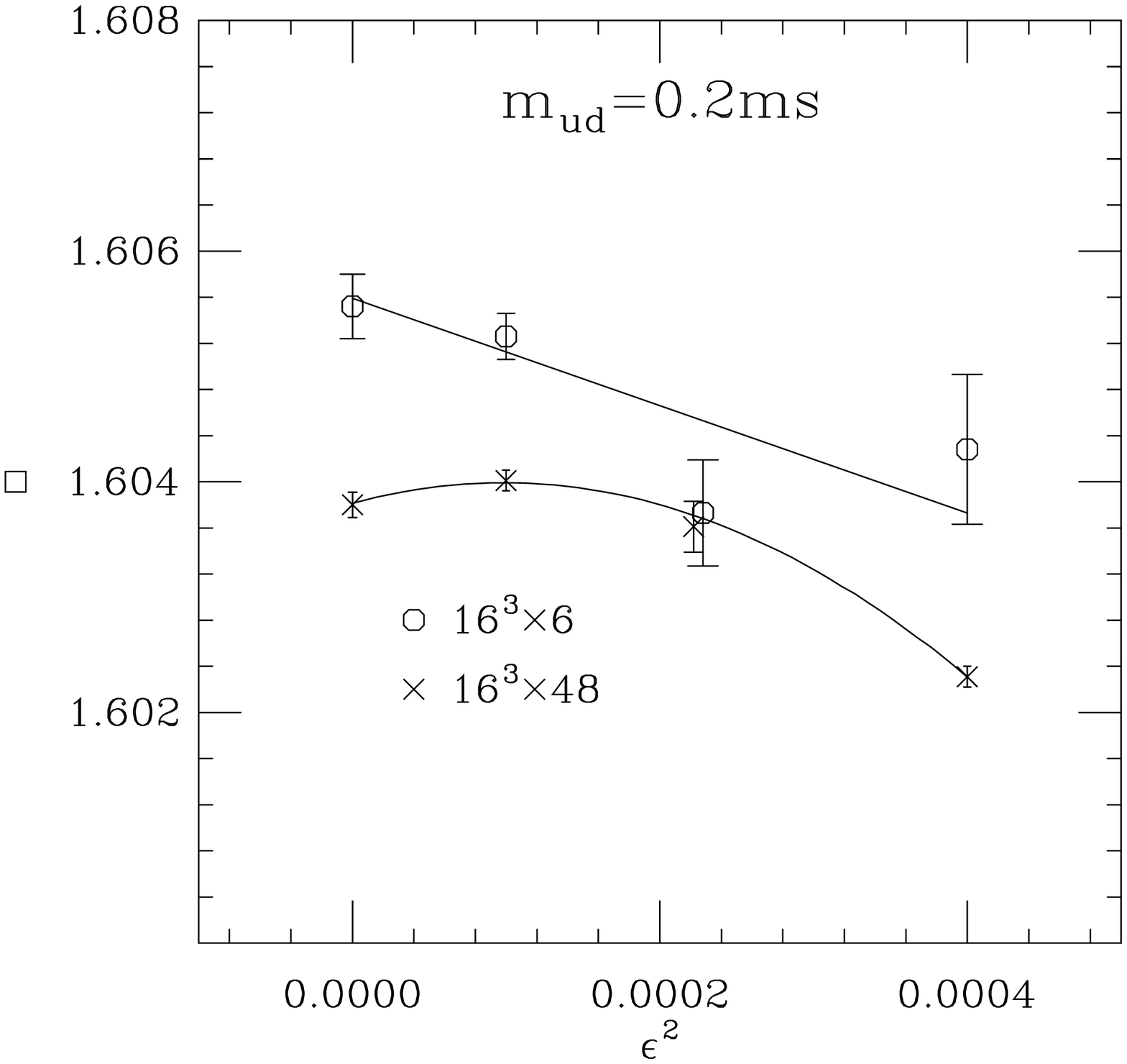}
&
  \epsfxsize=70mm
  \epsfbox{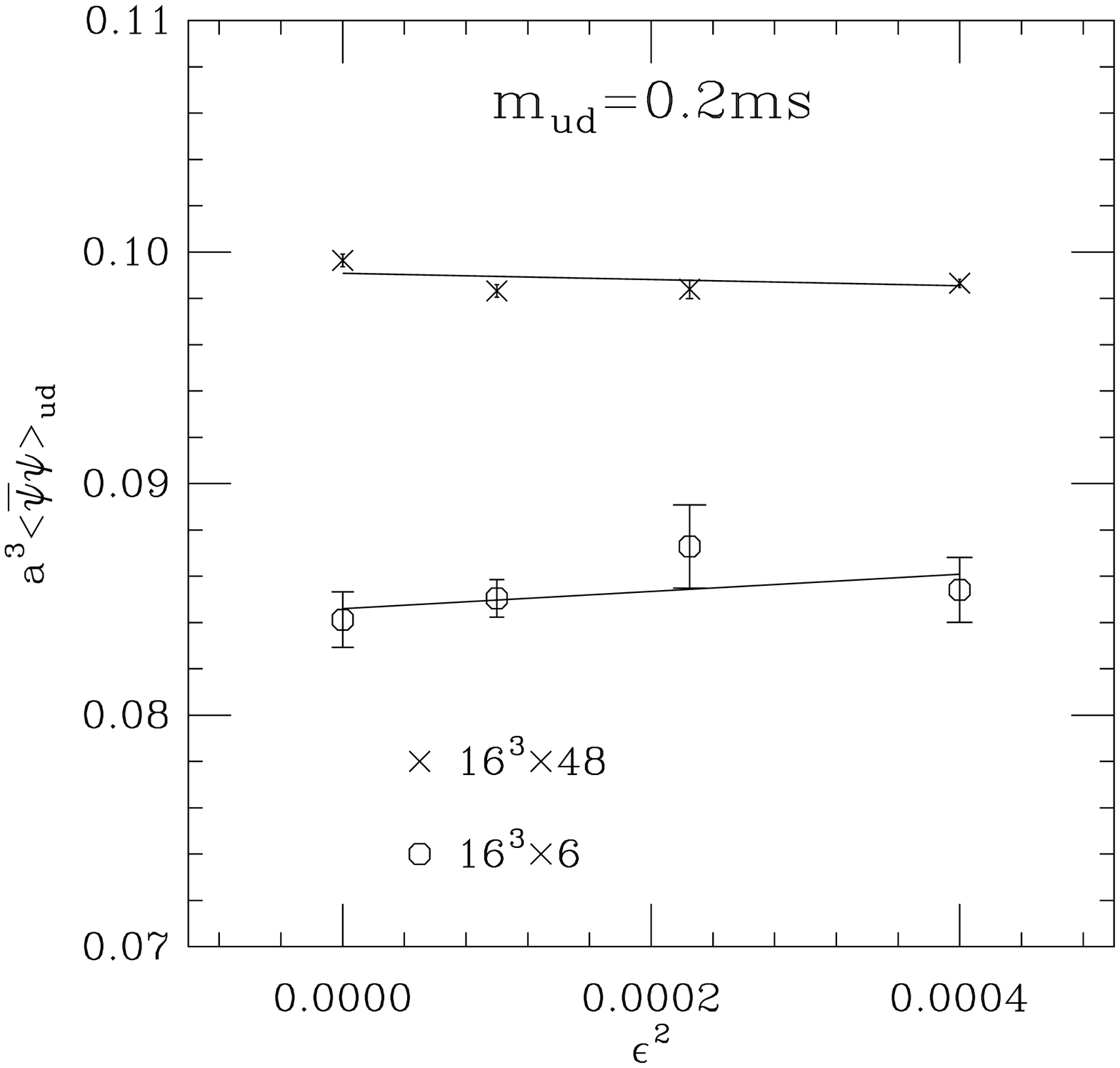}
\end{tabular}

\caption{Same as Fig.~\protect\ref{fig:obsvseps0.1ms} but for the
ensemble at $\beta = 6.467$, $am_{ud} = 0.01676$ and $am_s = 0.0821$.
The squared step size for the production of this ensemble is 0.0001 }
\label{fig:obsvseps0.2ms}
\end{figure}

The leapfrog-inspired $R$ algorithm is specifically designed to be a
second order integration algorithm.  That is, the truncation error at
the end of a molecular dynamics trajectory of fixed length decreases
with the integration step size $\epsilon$ as $\epsilon^2$.  In
Figs.~\ref{fig:obsvseps0.1ms} and \ref{fig:obsvseps0.2ms} we show the
step-size dependence of the plaquette and chiral condensate for the
improved action for one pair of ensembles.  On the larger-volume
zero-temperature lattices, we have found that the variation of the
plaquette with decreasing step size shows more apparent curvature over
this range of step sizes than do the smaller volume high temperature
lattices.  Consequently, as shown, we fit the low temperature results
to a quadratic in $\epsilon^2$.  Clearly both low and high temperature
values are subject to correction.  The corrections tend to cancel in
the difference. For the improved action the slopes for all seven
observables needed for the equation of state are tabulated in
Tables~\ref{tab:stepsizeplaq} and \ref{tab:stepsizepbp}.  For the high
temperature ensembles the slope is determined by a linear fit in
$\epsilon^2$.  For the zero temperature ensembles it is determined
from results at our smallest available pair of values of $\epsilon^2$,
treating an RHMC step size as 0, of course.

\begin{figure}
\begin{tabular}{ll}
  \epsfxsize=70mm
  \epsfbox{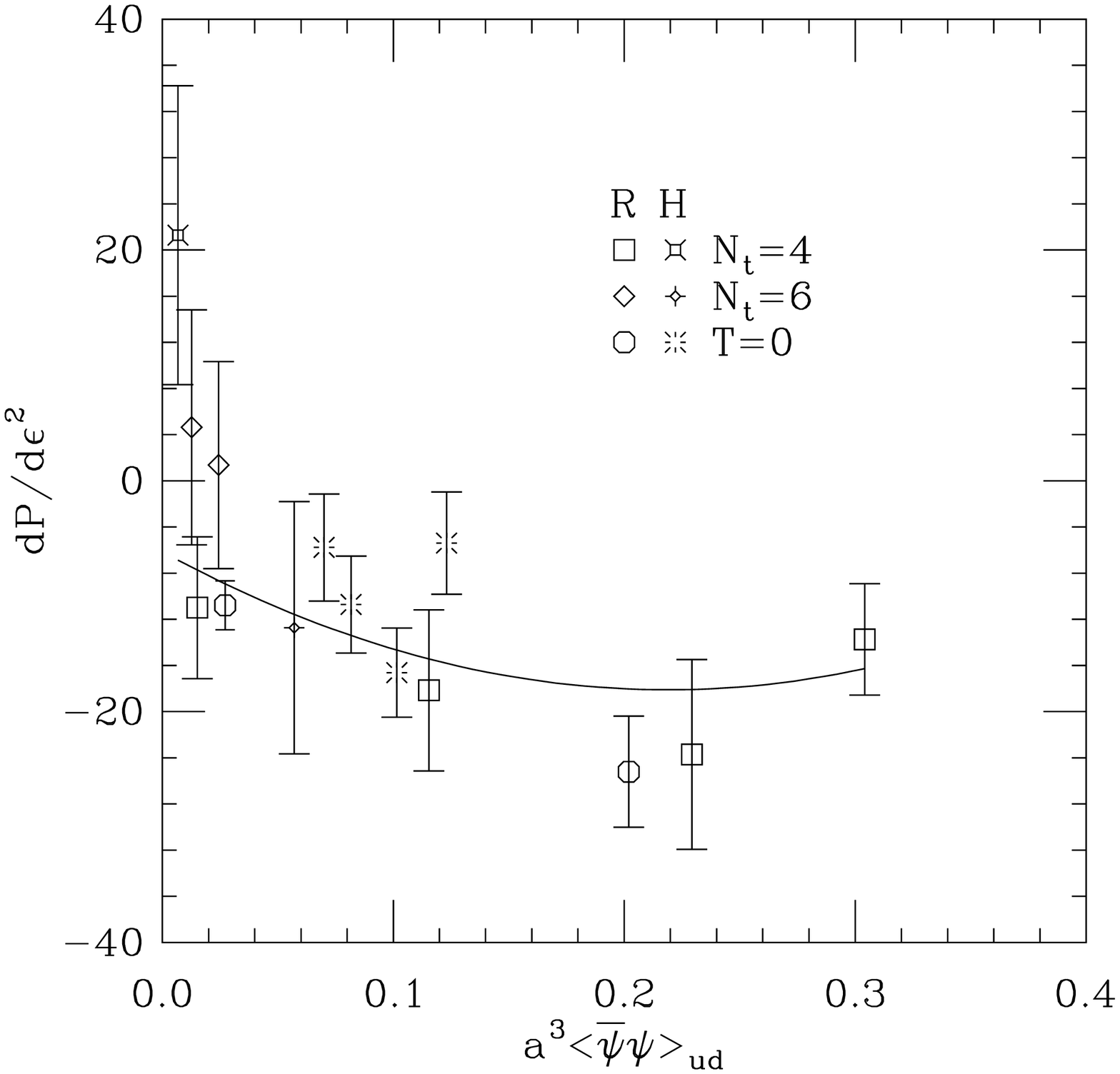}
&
  \epsfxsize=70mm
  \epsfbox{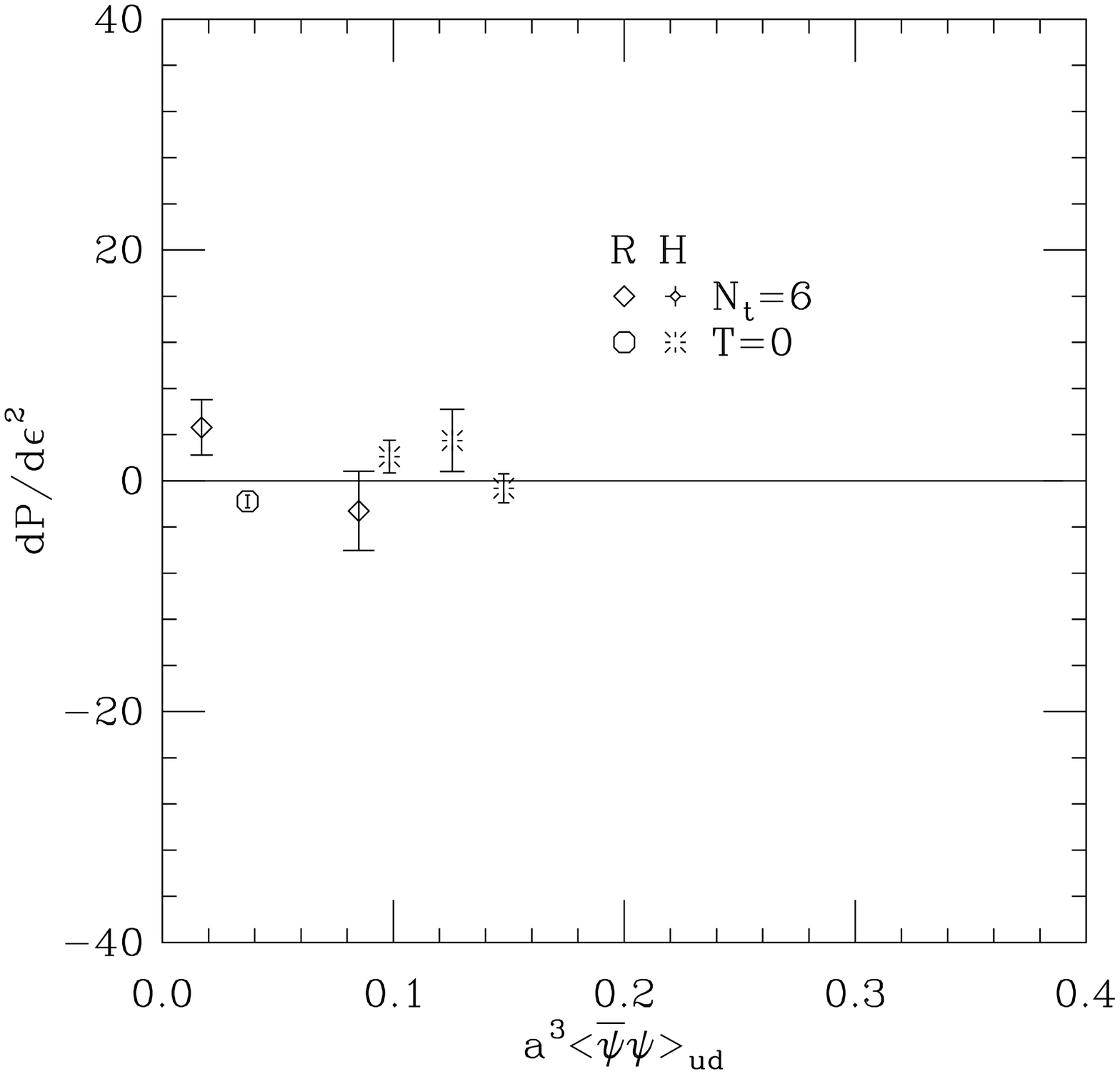}
\end{tabular}
\caption{Plaquette slopes for the $m_{ud}\approx 0.1\, m_s$ trajectory  
(left) and
the $m_{ud}\approx 0.2\, m_s$ trajectory (right) {\it vs.} the chiral  
condensate. On the left panel
  the quadratic fit to the data, Eq~(\ref{eq:stepsizeplaq}), is shown  
as well. Squares,
diamonds, and octagons indicate slopes determined from two different  
R algorithm step sizes.
Fancy squares, fancy diamonds, and bursts indicate slopes determined  
by comparing R-
with RHMC-algorithm measurements.
}
\label{fig:slope_pbp}
\end{figure}

The step-size correction depends largely on the size of the fermion
force, which is computed by inverting the fermion matrix.  Small
eigenvalues dominate the inverse.  The smallest eigenvalue is
controlled by the light quark mass.  Since the chiral condensate is
also determined from the inverse of the fermion matrix, we would
expect the chiral condensate and light quark mass to be natural
parameters for the step size error regardless of temperature.
Consequently, along a chosen trajectory of constant physics, we
parameterize the step-size slope of the plaquette for both high and
low temperature ensembles as a polynomial in the chiral condensate.
For the $m_{ud}\approx 0.1\, m_s$ trajectory we find it is modeled
reasonably well by the following quadratic form as shown in
Fig.~\ref{fig:slope_pbp}:
\begin{equation}
  \frac{dP}{d \epsilon^2} =  b(x-0.1)^2 + m(x-0.1) + c,
%RS   \frac{dP}{d \epsilon^2} =  a(p-0.1)^2 + m(p-0.1) + b,
\label{eq:stepsizeplaq}
\end{equation}
where $x = a^3 \pbp_{ud}$ is the light quark chiral condensate in  
lattice units.

The best fit values are
\begin{eqnarray}
b &=& 250(177), \ \ m = -109(54), \ \ c = -6.1(2.7),
%RSa &=& 291(180) \ \ m = -123(55) \ \ b = -5.7(2.7),
\label{eq:coeff}
\end{eqnarray}
for $\chi^2/df = 23/11$.  Thus our empirical model explains most of
the observed variation but not all.  We use it to estimate the
step-size correction along the $m_{ud}\approx 0.1\, m_s$ trajectory.  By
comparison the plaquette slopes for the $m_{ud}\approx 0.2\, m_s$
trajectory are small.  If we apply a correction according to a crude
linear fit to these slopes, the effect on the EOS is much smaller than
our statistical errors.  For these reasons we chose to ignore the
step-size error for this trajectory.

The fit to the step size correction also allows us to estimate the
error in our ability to predict the correction.  The largest error,
approximately 50\%, occurs at small values of the chiral condensate.
%RS occurs  --> occurs,
We take this as a conservative estimate of the error in our
correction throughout.

Table~\ref{tab:stepsizeplaq} shows, not surprisingly, that the
variation in all three gauge loop observables is correlated.  With our
normalization the slopes appear to be of comparable magnitude.  This
observation suggests generalizing the absolute plaquette correction to
all three gauge loops.

We also tabulate the slope of the fermion variables in
Table~\ref{tab:stepsizepbp}. Except for the gauge contribution,
all other estimated corrections to the EOS are negligible compared
with our statistical errors.  The gauge-action correction is   
smaller than
our statistical error in most cases; however, we have included it in the EOS
for the $m_{ud}\approx 0.1\, m_s$ trajectory since, although its effect is
comparable to the statistical error, it lowers all data points consistently.

\begin{table}[ht]
\begin{tabular}{lllllll}
\hline\hline
  volume           &$\beta$& $am_{ud}$ & $am_s $& $\VEV{P}$      &  $\VEV{R}$      & $\VEV{C}$\\
\hline
R $12^3 \times 4 $ &6.0    &0.0198   &0.1976 & $-$14(5)   &  $-$12(6)   &  $-$8(7)     \\
R $12^3 \times 4 $ &6.1    &0.0161   &0.1611 & $-$24(8)   &  $-$25(9)   &  $-$25(10)   \\
R $12^3 \times 4 $ &6.2    &0.0132    &0.1322 & $-$18(7)   &  $-$22(8)   &  $-$21(10)   \\
R $12^3 \times 12$ &6.2    &0.0132    &0.1322 & $-$25(5)   &  $-$26(5)   &  $-$26(7)    \\
H $16^3 \times 48$ &6.35   &0.00996   &0.0996 &  $-$5(4)   &   $-$5(6)   &  $-$5(7)	    \\
H $16^3 \times 48$ &6.35   &0.0206    &0.1001 & $-$0.6(1.3)&  $-$1.0(1.6)& $-$1.3(1.9)  \\
H $16^3 \times 48$ &6.40   &0.00909   &0.0909 & $-$17(4)   &  $-$20(5)   &  $-$17(4)    \\
H $16^3 \times 48$ &6.40   &0.01886   &0.0909 &  4(3)    &    5(3) &    5(4)    \\
R $12^3 \times 4 $ &6.458  &0.0082    &0.082 & $-$11(6)   &  $-$14(8)   &  $-$20(9)    \\
H $16^3 \times 6 $ &6.458  &0.0082   &0.082 & $-$18(4)   & $-$18(4)    &  $-$13(11)   \\
H $16^3 \times 48$ &6.458  &0.0082   &0.082 & $-$11(4)   & $-$11(7)    & $-$14(7)	    \\
H $16^3 \times 6 $ &6.467  &0.01676   &0.0821 & $-$3.8(1.6)&  $-$5(2)    &  $-$5(3)	    \\
H $16^3 \times 48$ &6.467  &0.01676   &0.0821 &  2.1(1.4)&   3(2) &  4(2)	    \\
H $16^3 \times 48$ &6.50   &0.00765   &0.0765 & $-$6(5)    &   $-$5(6)   &  $-$4(7)	    \\
R $12^3 \times 6 $ &6.55   &0.00705   &0.0705 & 1.4(9)   &  4(14)&  6(13)	    \\
R $12^3 \times 12$ &6.565  &0.005     &0.0484 & $-$36(4)   &  $-$46(6)   &  $-$48(6)    \\
R $12^3 \times 6 $ &6.65   &0.00599   &0.0599 &   5(10)  &  1(13)&   5(11)    \\
H $12^3 \times 4 $ &6.76   &0.005     &0.082 &  22(11)  &  19(18)&  23(16)    \\
R $24^3 \times 64$ &6.76   &0.005     &0.05   & $-$11(2)   &  $-$15(3)   & $-$10(3)     \\
R $20^3 \times 6 $ &6.76   &0.01      &0.05   &   5(2)   &    4(4)&   4(3)     \\
R $20^3 \times 64$ &6.76   &0.01      &0.05   & $-$1.8(6)  &  $-$2.9(4)  &  $-$2.6(6)   \\
R $20^3 \times 64$ &6.79   &0.02      &0.05   &  1.24(12)&   1.12(14)&   1.23(10) \\
R $20^3 \times 64$ &6.81   &0.03      &0.05   &  2.04(7) &   2.04(9)&   2.09(10)\\
\hline\hline
\end{tabular}
\caption{Step-size slope $d{\cal O}/d\epsilon^2$ for gauge field
contributions to the equation of state for a variety of lattice
ensembles. Three operators ${\cal O}$ are tabulated. 
The label R indicates values determined exclusively from  
the $R$ algorithm.
The label H indicates values determined with the aid of the RHMC  
algorithm.
}
\label{tab:stepsizeplaq}
\end{table}

\begin{table}[ht]
\begin{tabular}{llllll}
\hline\hline
  volume           & $am_{ud}$ &$\pbp_{ud}$&$\pbp_s$  & $\pbdmdup_{ud}$ & $\pbdmdup_s$\vspace{1.5mm}\\
\hline
R $12^3 \times 4 $ &0.0198  & 3(10)    &  4(5)     &  27(20)   &  45(19)   \\
R $12^3 \times 4 $ &0.0161  & 26(21)   &  19(11)   &  66(40)   &  70(37)   \\
R $12^3 \times 4 $ &0.0132  & 39(36)   &  26(16)   &  53(32)   &  54(29)   \\
R $12^3 \times 12$ &0.0132  & 15(11)   &  17(5)    &  44(17)   &  75(17)   \\
H $16^3 \times 48$ &0.00996  &$-$35(14)   &  $-$4(8)  & 12(16)  & 13(17)   \\
H $16^3 \times 48$ &0.0206   &$-$2(2)     &  $-$1(2)  &  1(5)   &  0(5)    \\
H $16^3 \times 48$ &0.00909  &$-$3(9)     & 3(41)     &  0(18)  &  0(18)   \\
H $16^3 \times 48$ &0.01886  & $-$15(5)   & $-$7(4)   &  7(7)   & 12(7)    \\
R $12^3 \times 4 $ &0.0082  & 1.6(1.4) &  9(6)       &  27(19)   &  45(19)  \\
H $16^3 \times 6 $ &0.0082  & 5(16)    & 10(8)       &  36(11)   &  44(12)  \\
H $16^3 \times 48$ &0.0082  &  $-$2(10)  &  3(5)     &   $-$     &  $-$     \\
H $16^3 \times 6 $ &0.01676  & $-$13(4)   &  $-$4(3)  &  11(2)    &  11(2)   \\
H $16^3 \times 48$ &0.01676  & 0.7(8)   &  1.8(7)     &   9(3)    &  7(3)    \\
H $16^3 \times 48$ &0.00765  & $-$5(10)   & $-$5(8)   & $-$3(15)  & 15(14)   \\
R $12^3 \times 6 $ &0.00705  & $-$20(25)  & $-$18(23) & $-$16(25) & $-$6(24) \\
R $12^3 \times 12$ &0.005    & $-$5(9)    & 18(6)     &  $-$      &  $-$     \\
R $12^3 \times 6 $ &0.00599  & 19(12)   & 33(20)    & 10(22)    & 5(30)      \\
H $12^3 \times 4 $ &0.005    &  0.2(4)  &  0.7(3)   & $-$30(28)   & $-$73(27) \\
R $24^3 \times 64$ &0.005    &  5.4(1.8)&  1.1(1.6) &  $-$6.9(2.9)& $-$2.7(3.3) \\
R $20^3 \times 6 $ &0.01     & $-$1.5(1.3)&  $-$2.7(2.7)&  $-$12(5) & $-$6(7) \\
R $20^3 \times 64$ &0.01     & $-$0.6(9)  &  1.2(6)   &  $-$  &  $-$	     \\
R $20^3 \times 64$ &0.02     & $-$1.1(3)  & $-$1.2(2)   &  $-$ &  $-$         \\
R $20^3 \times 64$ &0.03     & $-$0.66(11)& $-$0.61(9)  &  $-$ &  $-$\\
\hline\hline
\end{tabular}
\caption{Step-size slope $d{\cal O}/d\epsilon^2$ for fermion
contributions to the equation of state for the ensembles of Table
\protect\ref{tab:stepsizeplaq}.  Four operators ${\cal O}$ are
tabulated.}  \label{tab:stepsizepbp}
\end{table}

\end{document}